\documentclass[journal,10pt]{IEEEtran}
\usepackage{pifont,bm,multicol,amsfonts,amsmath,color,amssymb, epsfig,graphicx,cite,booktabs,makecell,psfrag,subfigure,algorithm,enumerate,stfloats,
algorithmicx,epstopdf,mathrsfs,algpseudocode,epstopdf,mathabx, xcolor, array, colortbl}
\usepackage{graphicx}
\usepackage{algorithmicx}
\usepackage{algpseudocode}
\usepackage{setspace}
\usepackage{amsmath}
\usepackage{amsthm}
\usepackage{booktabs}
\usepackage{multirow}
\usepackage{booktabs}
\usepackage{textcomp}

\newtheorem{definition}{\textit{\textbf{Definition}}}

\newtheorem{theorem}{\textit{\textbf{Theorem}}}

\algdef{SE}[DOWHILE]{Do}{doWhile}{\algorithmicdo}[1]{\algorithmicwhile\ #1}
\newcommand{\RNum}[1]{\uppercase\expandafter{\romannumeral #1\relax}}
\small
\IEEEoverridecommandlockouts

\begin{document}

\title{ \fontsize{21 pt}{\baselineskip}\selectfont Enhancing Resource Utilization of Non-terrestrial Networks Using Temporal Graph-based Deterministic Routing }

\author{Keyi Shi, Jingchao Wang, Hongyan Li,~\IEEEmembership{Member,~IEEE}, and Kan Wang
\thanks{K. Shi and H. Li are with the School of Communications Engineering, Xidian University, Xi'an 710071, P. R. China. (Email: kyshi10091@163.com; hyli@xidian.edu.cn).}
\thanks{J. Wang is with the Peng Cheng Laboratory, Shenzhen 518055, China (e-mail: wangjc61s@163.com).}
\thanks{K. Wang is with the School of Computer Science and Engineering, Xi’an University of Technology, Xi’an 710048, China. (Email: wangkan@xaut.edu.cn).}}


\maketitle

\begin{spacing}{1}
\begin{abstract}
Deterministic routing has emerged as a promising technology for future non-terrestrial networks (NTNs), offering the potential to enhance service performance and optimize resource utilization. However, the dynamic nature of network topology and resources poses challenges in establishing deterministic routing. These challenges encompass the intricacy of jointly scheduling transmission links and cycles, as well as the difficulty of maintaining stable end-to-end (E2E) routing paths. To tackle these challenges, our work introduces an efficient temporal graph-based deterministic routing strategy. Initially, we utilize a time-expanded graph (TEG) to represent the heterogeneous resources of an NTN in a time-slotted manner. With TEG, we meticulously define each necessary constraint and formulate the deterministic routing problem. Subsequently, we transform this nonlinear problem equivalently into solvable integer linear programming (ILP), providing a robust yet time-consuming performance upper bound. To address the considered problem with reduced complexity, we extend TEG by introducing virtual nodes and edges. This extension facilitates a uniform representation of heterogeneous network resources and traffic transmission requirements. Consequently, we propose a polynomial-time complexity algorithm, enabling the dynamic selection of optimal transmission links and cycles on a hop-by-hop basis. Simulation results validate that the proposed algorithm yields significant performance gains in traffic acceptance, justifying its additional complexity compared to existing routing strategies.
\end{abstract}
\end{spacing}


\IEEEpeerreviewmaketitle

\begin{spacing}{1}
\section{Introduction}
Non-terrestrial networks (NTNs) have emerged as a promising solution for global high-speed Internet access, thanks to their extensive coverage and robust bandwidth capabilities \cite{9275613}. The continuous progress in aerial and space technologies, coupled with reduced manufacturing and launch costs, has notably hastened the development of NTNs. This acceleration is exemplified by the rapid establishment of mega-constellations like Starlink, OneWeb, and Telesat, highlighting the growing significance of NTNs in the future connectivity landscape. {The 3rd Generation Partnership Project (3GPP)} is actively dedicated to evolving 5G systems to support NTNs. Since 2017, 3GPP has released a series of documents focusing on network architecture, system configuration, and radio access \cite{9579443}. Concurrently, {the Internet Engineering Task Force (IETF)} network working group diligently analyzes the requirements of satellite constellations in the future Internet. Their analysis identifies efficient routing as a key enabler to enhance service performance and resource utilization within NTNs \cite{lhan-problems-requirements-satellite-net-03}.

Nevertheless, designing efficient routing strategies for NTNs is challenging due to the dynamics of their network topologies and resources \cite{9750956}. Over the years, researchers have explored and implemented diverse routing strategies tailored to specific NTNs. One widely adopted approach is the shortest path routing algorithm (SPR), designed to facilitate routing for pre-defined remote sensing transmission missions \cite{icc2019pan}. SPR models the network topology as a static graph over the mission duration, identifying the end-to-end (E2E) path with the minimum delay or hops. However, this approach lacks adaptability to changing network conditions and traffic demands. The snapshot graph-based routing algorithm (STR) extends SPR by employing a series of static snapshots to model the time-varying network topology and calculates E2E routing in each snapshot \cite{634801}. However, STR may not be able to determine feasible routing paths within a single snapshot when contacts or resources are scarce. Another routing strategy, the contact graph routing algorithm (CGR), incorporates the caching-and-forwarding capability of satellites in routing decisions, enabling multi-hop transmissions in disruption-tolerant scenarios \cite{7060480}. However, CGR prioritizes establishing the earliest connected E2E routing, potentially compromising optimal delay performance. Furthermore, the aforementioned strategies base routing decisions on bandwidth requirements over a long time duration, without allowing for the precise specification of traffic transmission times. Consequently, micro-bursts occur frequently, leading to uncertain delays and congestion.

Deterministic routing holds considerable promise within NTNs, offering the potential to enhance service performance and optimize resource utilization \cite{9472798}. This technology facilitates precise scheduling of transmission links and cycles at each hop along the routing path, ensuring strict adherence to E2E delay and jitter requirements. Moreover, it enables dynamic allocation of network resources within each cycle on demand, thereby improving the overall resource utilization. Despite the commendable efforts of the {Institute of Electrical and Electronics Engineers (IEEE) Time-Sensitive Networking (TSN)} and the {Internet Engineering Task Force (IETF) Deterministic Networking (DetNet)} committees \cite{9456872}, the implementation of deterministic routing in NTNs remains challenging. This challenge stems from two primary factors: \romannumeral 1) the high complexity associated with solving integer linear programming (ILP) for joint routing and scheduling falls short of meeting real-time processing requirements, and  \romannumeral 2) the dynamic nature of network topology and resources complicates the identification of stable E2E routing paths.

In response to these challenges, we introduce a temporal graph-based strategy to efficiently address the problem.

\begin{itemize}
	\item Initially, we utilize a time-expanded graph (TEG) \cite{9072305} to represent the heterogeneous resources of an NTN in a time-slotted manner, including contact topology, link capacity, node storage, link delay, and storage delay.
	\item With TEG, we formulate the deterministic routing problem comprehensively, incorporating a set of crucial constraints. {Subsequently, we transform this nonlinear problem equivalently into a solvable ILP format. This transformation involves the linearization of cross-cycle propagation and caching constraints arising from long link delay and potential storage delay, thereby providing a robust yet time-consuming performance upper bound.}
 \item {To address the problem with reduced complexity, we construct an extended TEG (ETEG) to uniformly represent heterogeneous network resources and traffic transmission requirements. With ETEG, we propose a polynomial-time complexity algorithm for determining optimal transmission links and cycles on a hop-by-hop basis.}
\item {Furthermore, we analyze the optimality and complexity of the proposed algorithm, followed by an implementation framework based on segment routing to facilitate its feasibility within large-scale NTNs. Simulation results demonstrate the superior performance of our proposal over SPR, STR, and CGR in terms of traffic acceptance. Additionally, it exhibits a significantly lower running time than the ILP-based strategy (referred to as ILPS).}
\end{itemize}
\end{spacing}


\section{System model and problem formulation}

\begin{spacing}{1}
\subsection{System model}
In Fig. 1(a), we analyze an NTN, comprising  $N$ satellites denoted by the set $V = \left\{u_1,u_2,...,u_{N}\right\}$. These satellites are interconnected through time-varying yet predictable transmission links.  To enable precise delay control, the time window of each satellite is finely divided into consecutive cycles $\left\{\tau_h| h \ge 1 \right\}$ of equal duration $|\tau|$, where $\tau_h = \left ((h-1) \cdot |\tau|, h \cdot |\tau| \right]$. This division allows for treating the network topology as static within each cycle. Additionally, we consider a time-critical (TC) traffic demand, denoted as $f$, characterized by a period of $T_f$ and a per-period size of $A_f$. Assume that $f$ is injected into the NTN at time $t_f$ and needs to be delivered from a source satellite, $s \in V$, to a destination satellite, $d \in V$, within an upper bound of E2E delay, $B_f$. Then, we select the planning horizon, $D$, spanning from the start-cycle, $\tau_{\widecheck h}$, to the end-cycle, $\tau_{\widehat h}$, where $t_f$ and $t_f + B_f$ fall within $\tau_{\widecheck h}$ and $\tau_{\widehat h}$, respectively\footnote{To enable deterministic transmission in all traffic periods, we establish deterministic routing for $f$ in the first period and evolve it through repetition with cycle offset or necessary revisions.}. As illustrated in Fig. 1(c), we utilize a TEG, denoted as $\mathcal G = \left\{\mathcal V, \mathcal E, \mathcal C, \mathcal L\right\}$, to model heterogeneous network resources in a time-slotted manner. Specifically, $\mathcal G$ includes:

\begin{figure}[!t]	
	\centering{\includegraphics[width=0.95\linewidth]{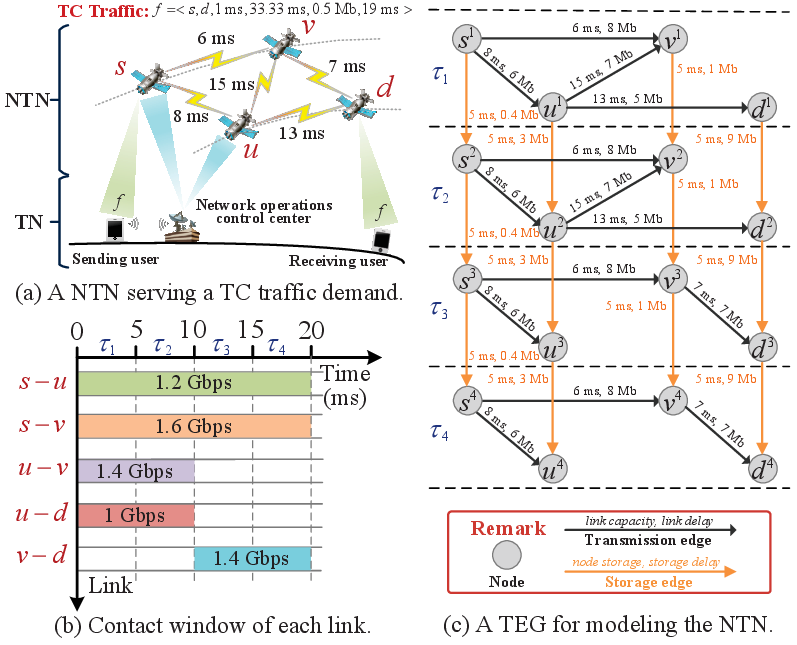}}
	\caption{Modeling a typical NTN using a TEG.}
	\label{fig:NTN_Sce2}
\end{figure}

\begin{itemize}
	\item A set of nodes, denoted as $\mathcal V = \left\{ u^h|u\in V, \widecheck h \le h \le {\widehat h}\right\}$, where each $u^{{h}}\in \mathcal
 V$ signifies a satellite $u$ within cycle $\tau_{h}$.
	\item A set of edges, denoted as $\mathcal E$, encompasses both transmission edges $\mathcal E_t$ and storage edges $\mathcal E_s$. Herein, ${\mathcal E_t} = \left\{ (u^{h},v^{h}) |u^{h},v^{h} \in \mathcal V, \widecheck h \le  h  \le \widehat h \right\}$ denotes the transmission links between satellites $u$ and $v$ in $\tau_{h}$. Additionally, ${\mathcal E_s}  =\left\{ (u^{h},u^{h+1})|u^{h}, u^{h+1} \in \mathcal V, \widecheck h \le  h  \le \widehat h - 1 \right\}$ depicts the capability of each satellite $u$ to cache data across adjacent cycles (e.g., from $\tau_{h}$ to $\tau_{h+1}$).
	\item A capacity set, denoted as $\mathcal C$, comprises two distinct subsets: a link capacity subset, $\mathcal C_t$, and a node storage subset, $\mathcal C_s$. Herein, ${\mathcal C_t} = \left\{ {c_{u^{h},v^{h}}}|(u^{h},v^{h}) \in {\mathcal E_t}, \widecheck h \le h \le \widehat h \right\} $ signifies the maximum amount of data that can be transmitted on each transmission edge  $(u^h,v^h)$, measured in megabytes (Mb). Furthermore, ${\mathcal C_s} = \left\{ {c_{u^{h},u^{h+ 1}}}|(u^{h},u^{h + 1}) \in {\mathcal E_s}, \widecheck h \!\le\! h \!\le\! \widehat h \!-\! 1\!\right\}$ represents the on-board storage resources of each satellite $u$ during any cycle $\tau_h$, measured in Mb.
	\item A delay set, denoted as $\mathcal L$, encompasses two distinct subsets: a link delay subset, $\mathcal L_t$, and and a storage delay subset, $\mathcal L_s$. Specifically, ${\mathcal L_t} = \left\{ {l_{u^{h},v^{h}}}|(u^{h},v^{h})  \in {\mathcal E_t}, \widecheck h \le h \le \widehat h\right\}$ represents the propagation delay on each transmission edge $(u^h,v^h)$, measured in milliseconds (ms). Additionally, ${\mathcal L_s} = \left\{ {l_{u^{h},u^{h + 1}}}|(u^{h},u^{h + 1}) \in {\mathcal E_s}, \widecheck h \le h \le \widehat h - 1 \right\}$ depicts the cross-cycle caching delay (e.g., from $\tau_h$ to $\tau_{h+1}$) at each satellite $u$, measured in ms. Without loss of generality, we can set $l_{u^h,u^{h+1}} = |\tau|$ for any $l_{u^h,u^{h+1}} \in \mathcal L_s$.
\end{itemize}



\subsection{Constraint establishment}
The deterministic routing problem focuses on effectively scheduling E2E transmission for the TC traffic demand $f$. This entails a judicious selection of satellites and links, along with the identification of suitable cycles for the transmission and caching of $f$ at each satellite. To facilitate a comprehensive problem formulation, we define binary-valued variables for all edges in $\mathcal G$, denoted as $X = \left\{x_{u,v}^{h,k}|(u^h,v^k) \in \mathcal E\right\}$, accounting for two distinct cases: for a transmission edge $(u^h,v^h)\in \mathcal E_t$, $x_{u,v}^{h,h} = 1$ indicates that $f$ is transmitted from satellite $u$ to satellite $v$ within the cycle $\tau_h$; otherwise, $x_{u,v}^{h,h} = 0$. Concerning a storage edge $(u^h,u^{h+1}) \in \mathcal E_s$, if $f$ is cached by satellite $u$ from cycle $\tau_h$ to cycle $\tau_{h+1}$, then $x_{u,u}^{h,h+1} = 1$; otherwise, $x_{u,u}^{h,h+1} = 0$.

\end{spacing}
\subsubsection{E2E transmission constraint} $f$ is required to initiate from the source satellite, $s$, and be transmitted to the destination satellite, $d$, within the planning horizon, $D$, i.e.,

\begin{spacing}{1}
{\small
\begin{align}
\sum\limits_{v^h:\left(s^h,v^{h}\right) \in \mathcal E_t}{\sum\limits_{h=\widecheck h}^{\widehat h}} {x_{s,v}^{h,h}} = \sum\limits_{{u^h}:\left(u^h,d^{h}\right) \in \mathcal E_t}{\sum\limits_{h={\widecheck h}}^{\widehat h}} {x_{u,d}^{h,h}} = 1.
\end{align}
}
\end{spacing}
\subsubsection{Lossless forwarding constraint} For any given satellite $v$ (excluding $s$ and $d$), it must forward the received $f$ within $D$, expressed as:
{\small
\begin{align}
\!\!\sum\limits_{u^h:\left(u^h,v^{h}\right) \in \mathcal E_t}\!{\sum\limits_{h={\widecheck h}}^{\widehat h}} {x_{u,v}^{h,h}} \! =\!\! \!\sum\limits_{{w^h}:\left(v^h,w^{h}\right) \in \mathcal E_t}\!{\sum\limits_{h={\widecheck h}}^{\widehat h}} {x_{v,w}^{h,h}},\! \forall v \! \in\! {V \setminus \left\{s,d\right\}}.
\end{align}
}
\subsubsection{Capacity constraint} For any transmission edge selected to transmit $f$, its link capacity should be no less than the traffic size $A_f$, expressed as:
\begin{spacing}{1}
\begin{align}
c_{u^{h},v^{h}} - x_{u,v}^{h,h} \cdot A_f \ge 0, \forall \left(u^{h},v^{h}\right) \in \mathcal E_t.
\end{align}
\end{spacing}

\subsubsection{Storage constraint} For any storage edge to cache $f$, its node storage should be no less than $A_f$, formulated as:
\begin{spacing}{1}
{\small
\begin{align}
c_{u^{h},u^{h+1}} - x_{u,u}^{h,h+1} \cdot A_f \ge 0, \forall \left(u^{h},u^{h+1}\right) \in \mathcal E_s.
\end{align}
}
\end{spacing}

{\subsubsection{Cross-cycle propagation and caching constraints}
When $f$ is transmitted from satellite $u$ to satellite $v$, the arrival time of $f$ at $v$ should occur later than the transmission cycle of $u$ but no later than the transmission cycle of $v$, i.e.,
\setcounter{equation}{5}
\small
\begin{align}
& \!\!\!\!\!O(u)  \!\!\cdot \! \!|\tau| \! - \!\!M \!\!\cdot\!(1 \!\!-\!y_{u,v}) \!\!<  \!\!y_{u,v}  \! \cdot  \!\left[\!L_t\!(u)  \!+  \!L_s\!(u) \!\! + \! \!t_f\!\right] \!\!+  \!l_{u,v}  \! \! \le \!\!O(v) \!\! \cdot \! \!|\tau| \tag{5a}.
\end{align}

\noindent Here,
\small
\begin{align}
O(u) = \sum\limits_{(u^h,w^h)\in \mathcal E_t}{\sum\limits_{h={\widecheck h}}^{\widehat h}}{h \cdot x_{u,w}^{h,h}} \tag{5b}
\end{align}
represents the transmission cycle\footnote{If $O(u) = 0$, it indicates that $u$ does not transmit $f$.} of $u$,
\small
\begin{align}
y_{u,v} = \sum\limits_{h={\widecheck h}}^{\widehat h} x_{u,v}^{h,h}, \tag{5c}
\end{align}
indicates whether $f$ is transmitted from $u$ to $v$,
{\small
\begin{align}
L_t(u) = \sum\limits_{(p^k,q^k)\in \mathcal E_t}\sum\limits_{k={\widecheck h}}^{O(u)-1}{x_{p,q}^{k,k}} \cdot l_{p^k,q^k} \tag{5d}
\end{align}
}represents the total propagation delay from $s$ to $u$,
\small
{
\begin{align}
L_s(u) = \sum\limits_{(p^k,p^{k+1})\in \mathcal E_s}\!\!\!\sum\limits_{k={\widecheck h}}^{O(u)-1} {x_{p,p}^{k,k+1}} \cdot l_{p^k,p^{k+1}} \tag{5e}
\end{align}}represents the total caching delay $s$ to $u$, together with the caching delay within $u$,
\small{
    \begin{align}
        l_{u,v} = \sum\limits_{k={\widecheck h}}^{\widehat h} x_{u,v}^{k,k} \cdot l_{u^k,v^k}, \tag{5f}
    \end{align}}represents the propagation delay from $u$ and $v$ within the transmission cycle of $u$, and $M$ represents a very large positive constant.
\subsubsection{Transmission timing constraint} The time when satellite $u$ sends $f$ satellite $u$ should fall within the transmission cycle of $u$, i.e.,
\small
\begin{align}
&\!\!\!\!\left[O(u) \!- \!1 \right] \!\!\cdot\!\! |\tau| \!-\! M \!\!\cdot \!\!(1\!\!-\!y_{u,v}) \!\!\le\! \!y_{u,v} \!\!\cdot \!\left[L_t(u) \!\!+\! \!L_s(u)\!\!+\!\!t_f\right] \!\!\le\!\! O(u)\! \!\cdot\! \!|\tau|, \tag{6}
\end{align}
\subsubsection{Caching timing constraint} When satellite $v$ needs to cache $f$ received from satellite $u$, the feasible cycles for caching should be no earlier that the cycle within which $f$ arrives at $v$ but earlier than the transmission cycle of $v$, expressed as:
\small
\begin{align}\label{Cache_timing1}
\!\!\!\!\!  y_{u,v}  \!\cdot\!  \left[L_t(u) \!+\!  L_s(u) \!+\!  t_f\right] \! +\! l_{u,v}\! - \!M\! \cdot \!(1\! \!- \! x_{v,v}^{h,h+1})\!\le\! h \!\cdot\! x_{v,v}^{h,h+1} \!\!\cdot \!|\tau|\! \nonumber \\ < \! O(v) \!\cdot \! |\tau|\! \!+\! \varepsilon \!\cdot \!(1 \!- \!x_{v,v}^{h,h+1\!}), \forall \widecheck h \!\le\! h \!\le \!\widehat h, \tag{7}
\end{align}
\noindent where $\varepsilon $ represents a very small positive constant.
\subsection{Problem formulation}
\setcounter{equation}{6}
\small
\begin{align}
&\!\!\!\!\mathbf {P_1 \! :} \min\limits_{X}\! \sum\limits_{(\!u^{h}\!,v^{h}\!) \in \mathcal E_t}\!{\sum\limits_{h={\widecheck h}}^{\widehat h}} x_{u,v}^{h,h}\! \cdot\!   {l_{u^{h}\!,v^{h}}} \!+\!\! \sum\limits_{(\!u^{h}\!,u^{h\!+\!1}\!) \in \mathcal E_s}\!{\sum\limits_{h={\widecheck h}}^{{\widehat h}-1}} x_{u,u}^{h,h\!+\!1} \!\!\cdot\!   {l_{u^{h}\!,u^{h\!+\!1}}} \tag{8}\\
&~~\rm{s.t.} (1)-(4),(\rm{5a}),(\rm{6}),(\rm{7}).\nonumber
\end{align}

The objective of $\mathbf{P_1}$ is to minimize the E2E delay, encompassing both the total propagation delay and caching delay during the delivery of $f$ from $s$ to $d$. Assuming the objective value is less than $B_f$, the solution to $\mathbf{P_1}$ provides efficient transmission scheduling for $f$ with deterministic guarantees. However, due to the presence of the term $y_{u,v} \!\cdot \!\left[L_t(u) \!+\!L_s(u)\!+\!t_f\right]$ in constraints (\rm{5a}), (\rm{6}), and (\rm{7}), where $L_t(u)$ and $L_s(u)$ involve variable-dependent summation and $y_{u,v}$ undergoes multivariable multiplication with both, these constraints become nonlinear. Consequently, $\mathbf{P_1}$ is unsolvable using existing ILP solvers. To address this challenge, we linearize these constraints by introducing auxiliary binary-valued variables along with a set of linear constraints.

Initially, we transform the variable-dependent summation in $L_t(u)$ and $L_s(u)$ into a summation term for the product of independent variables. This is achieved by introducing an auxiliary binary-valued variable, $\chi^u$, and the following linear constraints:
\setcounter{equation}{8}
\small
\begin{align}
&k - O(u)\! \ge \!(2\!-\!H) \!\cdot\! \chi^u + \varepsilon \!\cdot\! (1-\chi^u) - 1,  \forall {\widecheck h} \le k \le {\widehat h},
    \end{align}
\noindent and
\small
\begin{align}
&k - O(u)\! \le\! (H\!-\!1) \!\cdot \!(1 - \chi^u)\! -\! \varepsilon\! \cdot\! \chi^u, \forall {\widecheck h} \le k \le {\widehat h}.
    \end{align}
Here, $H \!=\! {\widehat h} \!-\! {\widecheck h} \!+\! 1$. Using $\chi^{u}$, $L_t(u)$ and $L_s(u)$ turn to be
{\small
\begin{align}
L_t^{(1)}(u) = \sum\limits_{(p^k,q^k)\in \mathcal E_t}\sum\limits_{k=\widecheck h}^{\widehat h}\chi^u \cdot {x_{p,q}^{k,k}} \cdot l_{p^k,q^k},
\end{align}}
\noindent and
\small
\begin{align}
L_s^{(1)}(u) = \sum\limits_{(p^k,p^{k+1})\in \mathcal E_s}\sum\limits_{k=\widecheck h}^{\widehat h -1} \chi^u \cdot {x_{p,p}^{k,k+1}} \cdot l_{p^k,p^{k+1}}.
\end{align}

Subsequently, we deal with multivariable multiplication in $y_{u,v} \cdot L_t^{(1)}(u)$ and $y_{u,v} \cdot L_t^{(1)}(u)$. For simplicity, we establish a general transformation paradigm for terms with the form $\prod\nolimits_{m=1}^{M}a_m$, where $a_m \in \left\{0,1\right\}$. Specifically, we introduce an auxiliary binary-valued variable, denoted as $\widetilde{a} =\prod\nolimits_{m=1}^{M}a_m $, for substitution, adhering to the following linear constraints:
{\small
\begin{align}
{\sum\nolimits_{m=1}^{M}a_m} - M + 1 \le \widetilde {a} \le a_m, \forall 1 \le m \le M.
\end{align}
}

Based on the paradigm, we can respectively transform $y_{u,v} \cdot L_t^{(1)}(u)$ and $y_{u,v} \cdot L_t^{(1)}(u)$ into
 \small
 \begin{align}
L_t^{(2)}(u)  = \sum\limits_{(p^k,q^k)\in \mathcal E_t}{\sum\limits_{h,k=\widecheck h}^{\widehat h}}\widetilde  {x}_{u,v,p,q}^{h,k} \cdot l_{p^k,q^k}
\end{align}
and
\small
  \begin{align}
L_s^{(2)}(u) = \sum\limits_{(p^k,p^{k+1})\in \mathcal E_s}{\sum\limits_{k=\widecheck h}^{\widehat h-1}} {\sum\limits_{h=\widecheck h}^{\widehat h}} \widetilde {x}_{u,v,p,p}^{h,k} \cdot l_{p^k,p^{k+1}},
\end{align}
with the introduced variables $\widetilde {x}_{u,v,p,p}^{h,k}$ and $\widetilde {x}_{u,v,p,p}^{h,k}$ satisfying
\small
\begin{align}
&x_{u,v}^{h,h} + {x_{p,q}^{k,k}} +
 \chi^u - 2 \le \widetilde  {x}_{u,v,p,q}^{h,k} \le x_{u,v}^{h,h},\\
&x_{u,v}^{h,h} + {x_{p,q}^{k,k}} +
 \chi^u - 2 \le \widetilde  {x}_{u,v,p,q}^{h,k} \le {x_{p,q}^{k,k}},\\
&x_{u,v}^{h,h} + {x_{p,q}^{k,k}} +
 \chi^u - 2 \le \widetilde  {x}_{u,v,p,q}^{h,k} \le \chi^u,\\
&x_{u,v}^{h,h} + {x_{p,p}^{k,k+1}} + \chi^u -2 \le \widetilde {x}_{u,v,p,p}^{h,k} \le x_{u,v}^{h,h},\\
&x_{u,v}^{h,h} + {x_{p,p}^{k,k+1}} + \chi^u -2 \le \widetilde {x}_{u,v,p,p}^{h,k} \le {x_{p,p}^{k,k+1}},\\
&x_{u,v}^{h,h} + {x_{p,p}^{k,k+1}} + \chi^u -2 \le \widetilde {x}_{u,v,p,p}^{h,k} \le \chi^u.
\end{align}

Substituting (\rm{14}) and (\rm{15}) into (\rm{5a}), (\rm{6}), and (\rm{7}), we can obtain the following linear constraints:
\small
\begin{align}
& \!\!\!\!\!\!O(u) \! \!\cdot \!\! |\tau| \!\!-\!\! M \!\!\cdot\!\! (1 \!\!- \!y_{u,v}) \!\!<\! \! L_t^{\!(2)}\!(u) \! + \!\! L_s^{\!(2)}\!(u)\! \!+\!\! y_{u,v} \! \!\cdot \!t_f \!+\! \! l_{u,v} \!\!\le\!\!  O(v) \!\!\cdot \!\! |\tau|,
\end{align}
\small
\begin{align}
&\!\!\!\!\!\!\left[ O(u) \!\!-\!\! 1 \right] \!\!\cdot\! \! |\tau| \!\!-\! \!M \!\!\cdot \!\!(1 \!\!- \!y_{u,v}) \!\!\le\!\! L_t^{\!(2)}\!(u)  \!+ \!\! L_s^{\!(2)}\!(u) \!+\! \!y_{u,v}  \!\cdot\! t_f \!\!\le\!\! O(u) \!\!\cdot \!\! |\tau|,
\end{align}
\noindent and
\small
\begin{align}\label{Cache_timing1}
\!\!\!\!\!\!  L_t^{\!(2)}(u) \!\!+\!\!  L_s^{\!(2)}(u) \!\!+\!  y_{u,v} \!\cdot \!t_f \! +\! l_{u,v}\!\! - \!M\! \cdot \!(1\! \!- \! x_{v,v}^{h,h+1})\!\le\! h \!\cdot\! x_{v,v}^{h,h+1} \!\!\cdot \!|\tau|\! \nonumber \\ < \! O(v) \!\cdot \! |\tau|\! \!+\! \varepsilon \!\cdot \!(1 \!- \!x_{v,v}^{h,h+1\!}), \forall \widecheck h \!\le\! h \!\le \!\widehat h.
\end{align}

Ultimately, $\mathbf{P_1}$ can be reformulated as an ILP problem:
\begin{spacing}{1}
{\small
\begin{align}
&\!\!\!\mathbf {P_2 \! :} \min\limits_{X'}\!\!\!\sum\limits_{(\!u^{h}\!,v^{h}\!) \in \mathcal E_t}\!{\sum\limits_{h=\widecheck h}^{\widehat h}} x_{u,v}^{h,h}\!\! \cdot\!   {l_{u^{h}\!,v^{h}}} \!\!+\!\!\! \sum\limits_{(\!u^{h}\!,u^{h\!+\!1}\!) \in \mathcal E_s}\!{\sum\limits_{h=\widecheck h}^{\widehat h\!-\!1}} x_{u,u}^{h,h\!+\!1} \!\!\cdot\!   {l_{u^{h}\!,u^{h\!+\!1}}} \tag{8} \\
&~~\rm{s.t.}~(\rm{1})-(\rm{4}), (\rm{9}),(\rm{10}), (\rm{16})-(\rm{24}),\nonumber
\end{align}
}
\end{spacing}
\noindent where the set of decision variables is defined as $X' = X \bigcup \left\{\chi^u|u\in V\right\}\bigcup\left\{\widetilde  {x}_{u,v,p,q}^{h,k}|(u^h,v^h),(p^k,q^k)\in \mathcal E_t \right\}\bigcup \left\{\widetilde {x}_{u,v,p,p}^{h,k}| \right. \\ \left.(u^h,v^h)\in \mathcal E_t,(p^k,p^{k+1})\in \mathcal E_s\right\}$.  {Notably, ILP solvers \cite{gurobi23} can effectively handle $\mathbf{P_2}$, providing a robust performance upper bound for $\mathbf{P_1}$. Nevertheless, the computations involved in these solvers prove excessively time-consuming, failing to meet real-time processing requirements. Therefore, we introduce a graph-based method to address $\mathbf{P_1}$, aiming to reduce the running time while ensuring optimality.}

\section{Temporal graph-based deterministic routing}
\subsection{Extended time-expanded graph model}
To effectively address $\mathbf{P_1}$, we enhance the original graph $\mathcal G$ to form an ETEG, denoted as $\mathcal G' = \left\{\mathcal V', \mathcal E', \mathcal C', \mathcal L'\right\}$, by introducing virtual nodes and edges to establish a uniform representation of heterogeneous network resources and traffic transmission requirements. The construction of $\mathcal G'$ is outlined as follows, as illustrated in Fig. 2:

\begin{figure}[!t]	
	\centering{\includegraphics[width=0.8\linewidth]{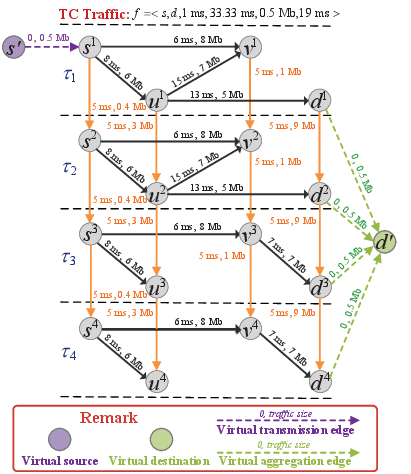}}
	\caption{An ETEG model.}
	\label{fig:ETEG_Modeling1}
\end{figure}

$\textit{1) }$To capture time-slotted network resources, we initialize the node set $\mathcal V'$ as $\mathcal V$, the edge set $\mathcal E'$ as $\mathcal E$, the capacity set $\mathcal C'$ as $\mathcal C$, and the delay set $\mathcal L'$ as $\mathcal L$.

$\textit{2) }$To signify the earliest cycle within which $f$ departs from the source satellite $s$, we introduce a virtual source $s'$ into $\mathcal V'$ and a virtual transmission edge $(s', s^{\widecheck h})$ into $\mathcal E'$.

$\textit{3) }$To represent potential cycles within which $f$ is transmitted to the destination satellite $d$, we introduce a virtual destination $d'$ into $\mathcal V'$ and a set of virtual aggregation edges $\{(d^h, d')|{\widecheck h} \le h \le {\widehat h}\}$ into $\mathcal E'$.
By designating $d'$ as the unique destination, we can avoid traversing potential routing to the destination satellite within each cycle (i.e., $d^h$, where $\widecheck h \le h \le \widehat h$), thus significantly reducing the computational complexity of deterministic routing.

$\textit{4) }$To indicate the capacity requirement of $f$, we set the capacity metrics of all introduced virtual transmission edges and virtual aggregation edges as $A_f$. Since these edges lack physical counterparts, we assign a delay metric of $0$ to them, thus not affecting the overall E2E delay.
\subsection{ETEG-based deterministic routing algorithm}
Based on ETEG, we identify that the deterministic routing problem is equivalent to a path-finding one, rather than directly solving the ILP in $\mathbf{P_2}$. Using this insight, we propose an ETEG-based deterministic routing algorithm with low complexity (as detailed in $\textbf{Algorithm 1}$). The proposed algorithm jointly utilizes both link capacity and node storage, facilitating cross-cycle propagation and caching of $f$. Consequently, it dynamically selects optimal links and cycles on a hop-by-hop basis to establish a time-featured path ($\textbf{\textit{Definition} 1}$) that minimizes the E2E delay while meeting resource requirements.}

\begin{definition}
	A time-featured path, denoted as $\mathcal P_f$, can be represented by a node sequence $
s' \!\to \cdotp\! \cdotp \!\cdotp \!\to\! u^{h} \!\to\! v^{k} \!\to \cdotp \!\cdotp \!\cdotp \!\to \!d'$ in the ETEG, adhering to the condition: if $u \ne v$, then $h < k$ and $c_{u^h,v^{h}} \ge A_f$; otherwise $k =h +1$ and $c_{u^h,v^{k}} \ge A_f$.
\end{definition}
\begin{algorithm}[h]
	\caption{ETEG-based deterministic routing algorithm}
	\begin{algorithmic}[1]
		\State \textbf{Input:} $\mathcal G' = \{\mathcal V',\mathcal E',\mathcal C',\mathcal L'\}$.
		
		\State $\mathcal P_f=\emptyset$, $p\left(u^{h}\right)=\emptyset$, $\forall u^{h}\in \mathcal V'$, $L\left(s'\right)= t_f $, $L\left(v^{k}\right)=+\infty$, $\forall v^{k} \in \mathcal V’ \setminus \{s'\}$, $Q=\mathcal V’$;
		
		\While {$ d' \in Q $}
		\State $u^{h}=\mathop{\arg\min}\limits_{v^{k}\in  Q}{L\left(v^{k}\right)}$;
		\State $Q \leftarrow Q \setminus \{u^{h}\}$;
		
		\For{$v^{k} \in Q:(u^h,v^k) \in \mathcal E'$}
				
		\If{$c_{u^{h},v^{k}} \ge A_f$}
		\State $r = \left\lceil {\frac{1}{{|\tau |}} \cdot \left( {{L\left(u^{h}\right) + {l_{u^{h},v^{k}}}}}\right) } \right\rceil $;
		\If{$L\left(u^{h}\right) + {l_{u^{h},v^{k}}} < L\left(v^{r}\right)$ }
		\State {$L\left(v^{r}\right) = L\left(u^{h}\right) + {l_{u^{h},v^{k}}}$};
		\State $p\left(v^{r}\right) = u^{h}$;
		\EndIf
		\EndIf
		
		\EndFor
		
		\EndWhile
		
		\If{$L\left(d'\right) \le t_f + B_f$}
		\State $u^{h} = d'$;
		\While {$u^{h} \neq \emptyset$}
		\State {$\mathcal P_f \leftarrow \mathcal P_f \bigcup \{u^{h}\}$};
		\State $u^{h} = p\left(u^{h}\right)$;
		\EndWhile
		\EndIf
		\State \textbf{Output:} $\mathcal P_f$ with deterministic guarantees.
	\end{algorithmic}
\end{algorithm}
In $\textbf{Algorithm 1}$, we execute the path-finding process based on $\mathcal G' = \{\mathcal V',\mathcal E',\mathcal C',\mathcal L'\}$. To facilitate this process, we define two essential parameters at each node $u^h \in \mathcal V'$: the pre-node $p\left(u^h\right)$, indicating the previous hop of $u^h$ in the time-featured path $\mathcal P_f$, and the node delay $L\left(u^h\right)$, representing the propagation delay and caching delay along $\mathcal P_f$ from  $s'$ to $u^h$. Additionally, we introduce a priority queue, denoted as $Q$, for maintaining nodes awaiting the determination of their node delay. During initialization (in step 2), we set $\mathcal P_f = \emptyset$, $p\left(u^h\right) = \emptyset$ for any node $u^h \in \mathcal V'$, $L\left( s'\right) = t_f$, $L\left(v^k\right) = +\infty$ for any node $v^k \in \mathcal V' \setminus \{s'\}$, and $Q = \mathcal V'$. In each iteration (from steps 3 to 15), we extract $u^{h}$ with the minimum node delay from $Q$. Subsequently, we update each node $v^{k}$ adjacent to $u^h$, provided that the resources (i.e., link capacity or node storage) are sufficient, and the node delay of $v^{k}$ can be reduced via the relay by $u^h$. Due to cross-cycle propagation and caching constraints, the updated node might be $v^r$ with $r = \left\lceil \frac{1}{|\tau|} \cdot\left(L\left(u^h\right)+l_{u^h,v^k}\right)\right\rceil$. Consequently, we have $L\left(v^{r}\right) = L\left(u^{h}\right) + {l_{u^{h},v^{k}}}$ and designate $p\left(v^r\right) = u^h$. The above iteration continues until $d'$ is extracted from $Q$. Finally, if $L\left(d'\right) \le t_f + B_f $ holds, we can obtain $\mathcal P_f$ by backtracking from  $d'$ to $s'$ (from steps 17 to 21); otherwise, no feasible $\mathcal P_f$ exists.

Fig. 3 illustrates an application. Initiated from $s'$, the proposed algorithm updates the node delay of its sole neighbor, $s^1$, to $L\left(s^1\right) = 1~\rm{ms}$ and set $p(s^1) = s'$. Next, we pop $s^1$ from $\mathcal Q$ and traverse all its neighbors, yielding the following: $L\left(s^2\right) = +\infty$, due to insufficient node storage of $(s^1,s^2)$; $L\left(u^2\right)=9 \,\rm{ms}$ and $L\left(v^2\right)=7\,\rm{ms}$, owing to cross-cycle propagation from $s^1$ to $u^2$ and $v^2$, respectively. In iteration 3, $v^2$ is extracted from $Q$, and $L\left(v^3\right)$ is updated to $7+5=12\,\rm{ms}$ through cross-cycle caching from  $v^2$ to $v^3$. The proposed algorithm then selects $u^2$ and update its sole neighbor $u^3$ with $L\left(u^3\right) = 9 + 5 = 14 \,\rm{ms}$. Subsequent steps involve updating $L\left(d^4\right)$ to $12+7=19 \,\rm{ms}$ through cross-cycle propagation from $v^3$ to $d^4$, which also serves as the ultimate node delay at $d'$, i.e., $L\left(d'\right) = 19 \,\rm{ms} < 1 + 19 = 20 \,\rm{ms}$. Through backtracking, a feasible time-featured path $\mathcal P_f\!: \!s' \!\to\! s^1 \!\to\! v^2 \!\to\! v^3 \!\to\! d^4 \!\to\! d'$ is obtained.

\begin{figure}[!t]	
	\centering{\includegraphics[width=0.9\linewidth]{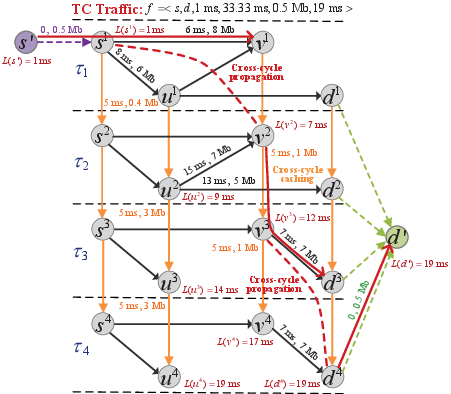}}
	\caption{An application of the ETEG-based deterministic routing algorithm.}
	\label{fig:Application1}
\end{figure}
\subsection{Complexity and optimality analyses}
\begin{theorem}
	The ETEG-based deterministic routing algorithm is capable of calculating a time-featured path with minimum E2E delay.
\end{theorem}
\begin{spacing}{1}
\begin{proof}
For $\textbf{Algorithm 1}$, we demonstrate that each node extracted from $Q$ has been determined with the minimum node delay. This assertion remains valid for $s^1$ with $L\left(s^1\right) = t_f$. Moving on to the $M$-th extracted node, denoted as $u^h$, with a node delay $L\left(u^h\right)$, we evaluate whether $L\left(u^h\right)$ can be further reduced through relaying by any node in $Q$, such as $w^r$. If so, either $L\left(w^r\right) + l_{w^r,u^r} < L\left(u^h\right)$ holds for cross-cycle propagation, or $L\left(w^r\right) + l_{w^r,w^{r+1}} < L\left(w^{r+1}\right) = L\left(u^h\right)$ holds for cross-cycle caching. However, due to step 4, $L\left(w^r\right) \ge L\left(u^h\right)$ holds, along with $l_{w^r,u^r} \ge 0$ and $l_{w^r,w^{r+1}} \ge 0$, thereby contradicting the aforementioned inequalities. Conversely, $L\left(u^h\right)$ is minimized when $u^h$ is extracted from $Q$, also holding for $d'$. The proof is completed.
\end{proof}
\end{spacing}
\begin{theorem}
The time complexity of the ETEG-based deterministic routing algorithm is $O\left(|\mathcal E'| \cdot \log |\mathcal V'|\right)$, where $|\mathcal V'|$ and $|\mathcal E'|$ denote the number of nodes and edges in the input $\mathcal G'$, respectively.
\end{theorem}
\begin{spacing}{1}
\begin{proof}
For $\textbf{Algorithm 1}$, assume that the $\mathcal G'$ and $Q$ are stored in adjacency lists and binary heaps, respectively. The initialization in step 2 takes $O\left(|\mathcal V'|\right)$ time. During each iteration from steps 3 to 15, it requires $O\left(1\right)$ time to extract $u^{h}$ with the minimum node delay from $Q$ and $O\left(\log |\mathcal V'|\right)$ time to update $Q$. Furthermore, updating all nodes adjacent to $u^{h}$ takes at most $O\left({\rm {deg}}\left(u^{h}\right) \cdot \log |\mathcal V'|\right)$, where ${\rm {deg}}\left(u^{h}\right)$ represents the out-degree of $u^{h}$ in $\mathcal G'$. At worst, we must traverse all nodes in $\mathcal V'$ once before extracting $d'$ from $Q$. Therefore, the time complexity reaches $O\left(|\mathcal V'| \cdot \left(1 + \log |\mathcal V'| + {\rm {deg}}\left(u^{h}\right)\cdot \log |\mathcal V'|\right) \right)= O\left(\left(|\mathcal V'|+|\mathcal E'|\right)\cdot \log |\mathcal V'|\right) $. Additionally, the backtracking process takes at most $O\left(|\mathcal V'|\right)$. Thus, the total time complexity is $O\left(|\mathcal V'|\right) + O\left(\left(|\mathcal V'|+|\mathcal E'|\right)\cdot \log |\mathcal V'|\right) + O\left(|\mathcal V'|\right) = O\left(|\mathcal E'| \cdot \log |\mathcal V'|\right)$, as $|\mathcal E'| \ge |\mathcal V'| - 1$ for a connected graph $\mathcal G'$. The proof is completed.
\end{proof}
\end{spacing}
\subsection{Algorithm implementation}
\begin{spacing}{1}
{Following \cite{SoftMulti}, we propose an implementation framework based on segment routing \cite{SegRout} for our algorithm. Within the NTN scenario in Fig. 1(a), we present the key aspects of the framework as follows:

\subsubsection{Parameter maintenance}
The network operations control center (NOCC) continuously acquires link status information from satellites through low propagation delay satellite-to-ground links. It extracts essential network parameters for deterministic routing decisions, including link capacity, link delay, and node storage.
\subsubsection{Routing decision}
Leveraging the maintained network parameters and traffic information from the terrestrial sending user, the NOCC constructs the ETEG and determines optimal deterministic routing for the TC traffic demand. Additionally, the NOCC not only reserves the required resources for the demand by updating network parameters but also configures the deterministic forwarding table sent to the specified source satellite.
\subsubsection{Routing deployment}
Following the deterministic forwarding table, the source satellite modifies TC traffic demand packets injected by the sending user. This modification involves encapsulating per-hop transmission link and cycle information into the packets' headers, directing them across the network transmission until reaching the destination satellite. Then, the packets are decapsulated and downlinked to the terrestrial receiving user.
\subsubsection{Routing evolution}
The NOCC consistently evaluates the feasibility of preceding deterministic routing for upcoming traffic periods. If feasible, the source satellite simply introduces a period-size cycle offset when deploying the deterministic forwarding table; otherwise, the NOCC re-executes the routing decision in \textit{2)} and the routing deployment in \textit{3)}.

Notably, a cross-domain decision architecture \cite{CrossDo} can be alternatively deployed when a single NOCC is insufficient to handle all TC traffic demands across the NTN, or when long propagation delays emerge as a primary concern.}
\end{spacing}

\section{Simulations}
\subsection{Simulation setup}
We conduct simulations on a partial Starlink constellation comprising 168 satellites selected from S1 \cite{vtieee45}. These satellites are distributed across 12 orbits, each accommodating 14 satellites, positioned at a height of 550 \rm{km} with an inclination of $53 ^\circ$. Employing the Satellite Toolkit (STK) simulator, we generate a time-varying NTN scenario with parameters in Table \RNum{1}. Consider that TC traffic demands continuously enter the NTN within the initial 120 seconds (s), following a Poisson process. The source-destination satellite pairs of these demands are randomly specified across the constellation. Each demand, representing applications like high-quality video telephony \cite{voip555}, operates with a period of 33.33 \rm{ms} (equivalent to a video frame rate of 30 frames per second) and has an active duration varying from 60 to 180 \rm{s}. Furthermore, the per-cycle size of each demand follows a uniform distribution between 0.05 \rm{Mb} and 0.6 \rm{Mb}, with an E2E delay upper bound set at 75 \rm{ms}.
Simulations are executed on a Hewlett-Packard Z620 tower workstation (Intel Core i9-13900H CPU@2.60GHz, 32G RAM, Windows 11 x64) with a C++ environment.
\begin{table}[!htbp] \scriptsize
	\caption{Network Parameters}\label{Tab.1}
 \centering
\begin{tabular}{|@{\hspace{0.5em}}c@{\hspace{0.5em}}|@{\hspace{0.5em}}c@{\hspace{0.5em}}|@{\hspace{0.5em}}c@{\hspace{0.5em}}|@{\hspace{0.5em}}c@{\hspace{0.5em}}|@{}c@{\hspace{0.5em}}|@{\hspace{0.5em}}c@{\hspace{0.5em}}|}
		\hline
Cycle duration (ms) &  Link capacity (Mb) & Node storage (Mb) & Link delay (ms)\\
\hline
$|\tau| =5$ & $c_{u^h,v^h} = 5$ & $c_{u^h,u^{h+1}} = 1000$ & $l_{v^h,v^h} \in \left[5,12\right]$\\
\hline
	\end{tabular}
\end{table}
\begin{spacing}{1}
{Throughout the entire 300-second horizon, we evaluate the performance of our proposed algorithm (referred to as DetR) and four benchmark strategies: SPR, STR, CGR, and ILPS, using the following metrics:}
\begin{spacing}{1}
\begin{itemize}
	\item \textit{Traffic acceptance} $\alpha$: The total size of demands with E2E deterministic transmission guarantees, providing insights into network resource utilization.

 	\item \textit{Average running time} $\beta$:  The average time to process a single demand.
	
	\item \textit{Average E2E path delay} $\gamma$: The average delay of routing paths associated with demands possessing E2E deterministic transmission guarantees.

\end{itemize}
\end{spacing}

\subsection{Simulation results}


\begin{figure}[!t]	
	\centering{\includegraphics[width=1\linewidth]{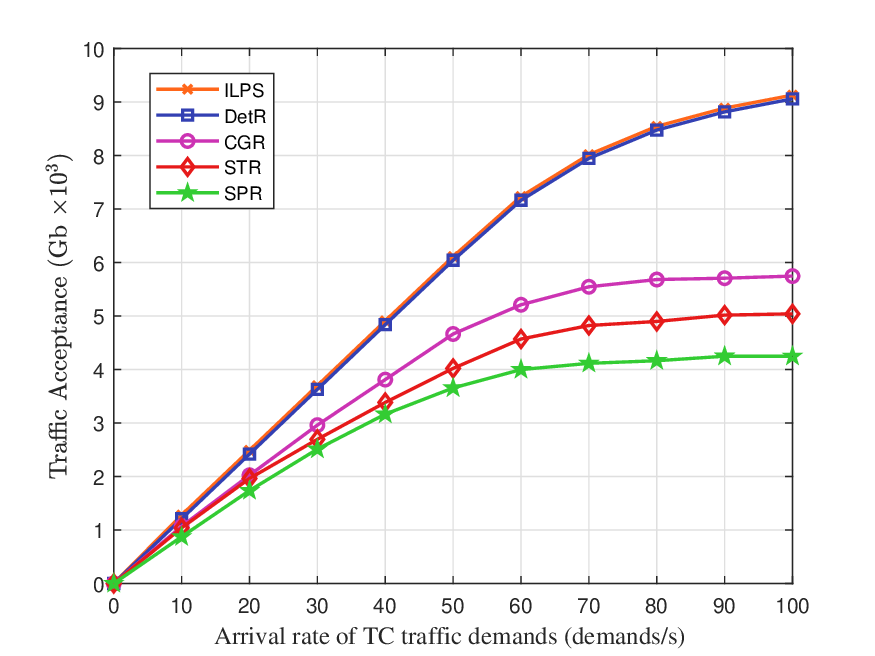}}
	\caption{Evaluation of traffic acceptance.}
	\label{fig:ETEG_Modeling1}
\end{figure}
We first evaluate the traffic acceptance, $\alpha$, by varying the arrival rate from 1 to 100 demands per second (demands/s), as shown in Fig. 4. As expected, $\alpha$ values for all algorithms increase as the arrival rate increases, but the increase rate gradually slows down. This happens as the increasing demands occupy most of the available network resources. Notably, DetR is comparable to the optimal ILPS and surpasses CGR, STR, and SPR. In particular, when the arrival rate reaches 100 demands/s, $\alpha$ is improved by more than 50\%. This significant enhancement is attributed to the ability of DetR and ILPS to jointly utilize link capacity and node storage in different cycles, enabling conflict-free routing paths for a larger number of demands. In contrast, SPR records the lowest performance due to its neglect of time-varying network characteristics, focusing solely on routing paths within a static graph. Consequently, these paths may encounter interruptions and congestion. Notably, STR employs a series of time-evolving snapshots to model the NTN, while CGR introduces connectivity between adjacent snapshots, expanding the solution space compared to SPR. However, since the routing paths are determined based on average bandwidth requirements, micro-bursts occur frequently among demands, limiting traffic acceptance.

\begin{figure}[!t]	
	\centering{\includegraphics[width=1\linewidth]{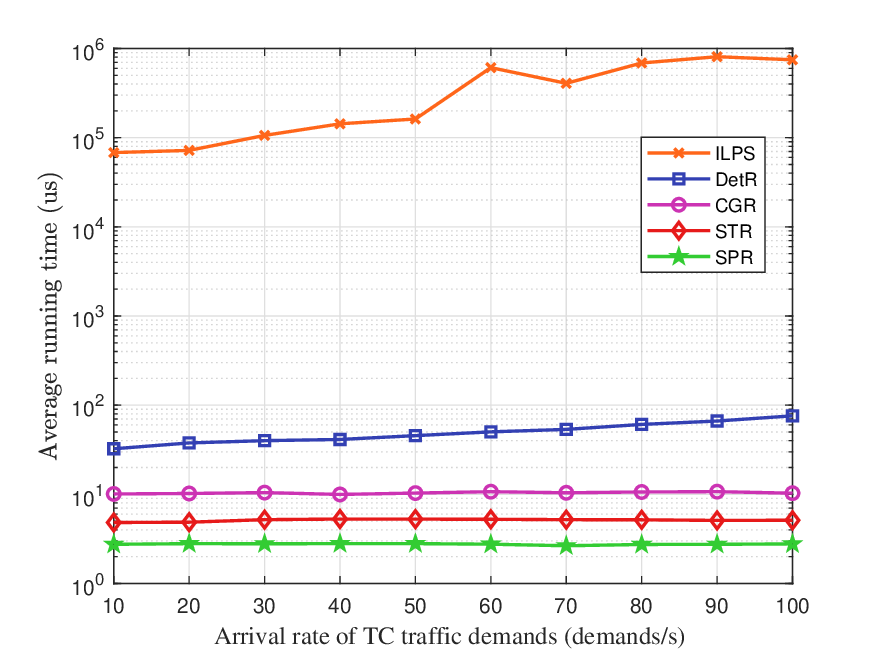}}
	\caption{Evaluation of average running time.}
	\label{fig:ETEG_Modeling1}
\end{figure}

{Fig. 5 illustrates the average running time, $\beta$, under varying arrival rates of demands. Notably, $\beta$  values for four graph-based algorithms are significantly lower than that of ILPS, primarily attributed to not traversing the entire solution space for routing decisions. DetR is the slowest among the four, with the gap not exceeding 80 microseconds (us). The increased complexity arises because DetR determines optimal transmission links and cycles for demands within a time-expanded solution space. Together with Fig. 4, it becomes evident that the commendable enhancement achieved by DetR in traffic acceptance justifies its increased complexity compared to SPR, STR, and CGR. Furthermore, DetR’s $\beta$ exhibits a gradual increase with rising arrival rates since DetR facilitates deterministic routing evolution by introducing cycle offsets or performing re-execution. These actions become more frequent as demands increase, thus increasing $\beta$.}


\begin{figure}[!t]	
	\centering{\includegraphics[width=1\linewidth]{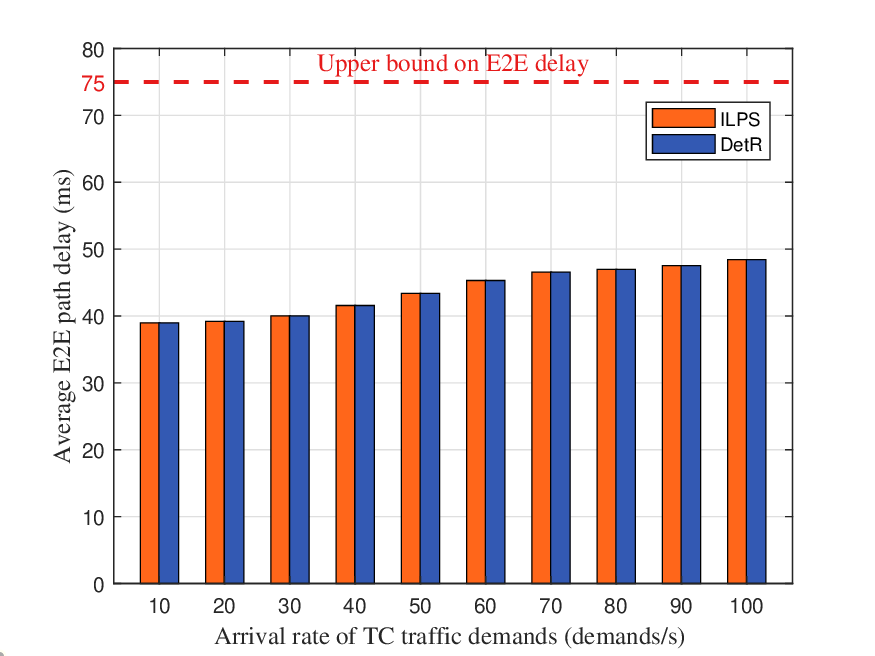}}
	\caption{Evaluation of average E2E path delay.}
	\label{fig:ETEG_Modeling1}
\end{figure}

{We also evaluate the average E2E path delay, $\gamma$, for both DetR and ILPS. As shown in Fig. 6, the $\gamma$ of DetR aligns with that of ILPS, gradually increasing as the arrival rate of demands rises. This trend is expected, as network resources become limited under heavy demand loads. To accommodate more demands while maintaining conflict-free deterministic routing paths, essential cross-cycle propagation and caching are included in these paths.  Although $\gamma$ increases, it remains below the upper bound of 75 \rm{ms}. Notably, the average E2E path delay of CGR, STR, and SPR is not presented, since their traffic acceptance is far lower than that of DetR, enabling them to find routing paths with lower delay under lightly loaded network conditions. Consequently, direct numerical comparisons among all algorithms lack meaningful insight.}

\section{Conclusion}
This study focuses on addressing the deterministic routing problem within NTNs. Leveraging the TEG, we meticulously formulated and equivalently transformed this intricate problem into a solvable ILP problem, providing a robust yet time-consuming performance upper bound. To enhance time efficiency, we introduced an ETEG-based deterministic routing algorithm with polynomial time complexity. This algorithm enables the joint utilization of link capacity and node resources, facilitating cross-cycle propagation and caching of the TC traffic demands. Consequently, it can determine optimal transmission links and cycles on a hop-by-hop basis. Simulation results demonstrated that our proposal outperforms SPR, STR, and CGR in terms of traffic acceptance, thereby justifying its additional complexity. Furthermore, it exhibits significantly reduced running time compared to ILPS.
\end{spacing}







\begin{spacing}{1}

\end{spacing}
%


\end{document}